# Comment on the Characterization of Local Structure in an Inhomogeneous Liquid

Linsey Nowack and Stuart A. Rice*

*James Franck Institute, Department of Chemistry, and the Chicago Center for Theoretical Chemistry, The University of Chicago, Chicago, IL, 60637*

*Corresponding author: sarice@uchicago.edu

## Abstract

The search for local structures within a disordered medium has led to proposals of several methods for probing transient short-range symmetry in a homogeneous monoatomic liquid. We offer a comparison of different characterizations of such local structure in an inhomogeneous liquid that has a density gradient. Here, we simulate the interfaces between a Lennard-Jones liquid and: (i) a flat Lennard-Jones wall, (ii) a wall of face-centered cubic-packed Lennard-Jones particles, and (iii) a Lennard-Jones vapor. The interfacial density distributions in cases (i) and (ii) have oscillations extending several particle diameters into the bulk liquid, while that in case (iii) the interfacial density distribution monotonically decays from the bulk liquid to the vapor. We search for transient ordered fluctuations, which we identify with local ordered structures, using the Aperture Cross Correlation Function (ACCF) of the system. We find that only particles in the FCC-liquid interface that are in the two layers closest to the FCC surface exhibit preferred four-fold symmetry. Furthermore, the absence of three- or five-fold symmetry peaks in the ACCF of the liquid-vapor interface yields a different picture of that interface than derived from the high density of bipyramidal clusters found in Royall et al's study (*Molecular Physics*, **109** (7-10), 1393-1402 (2011)). While the ACCF results need not be viewed as contradictory with Royall's Topological Cluster Classification (TCC) results, we argue the advantages of using the ACCF to characterize transient ordered fluctuations as being both experimentally accessible in favorable cases and not reliant on a pre-assumed list of possible structures.

## 1. Introduction

This short paper is concerned with adding a different perspective to the collection of current representations of the local structure in an inhomogeneous liquid. The inhomogeneous liquids that we are interested in are found in the liquid-vapor interface, the liquid-wall interface, and liquids confined in slits. The issue addressed is the interpretation of the choice of representation of local structure and the relationship of that choice to an experimental observable. Specifically, we propose using a variant of cross-correlation diffraction, sometimes called fluctuation diffraction, that can uniquely identify a transient local structure if it exists. Recent experimental and theoretical developments make it possible, in favorable cases, to generate cross correlation diffraction patterns in laboratory experiments [1-9].

We start our discussion by considering a homogeneous liquid. The suggestion that one can identify local structural motifs in a homogeneous liquid, e.g. clusters with well-defined symmetries, and that it is useful to do so with respect to interpreting various properties of the liquid, has a long history. The theme advanced by this suggestion is that subsets of particles in the liquid tend to form locally favored structures that have free energies that are lower, and lifetimes that are longer, than does a disordered particle distribution occupying the same volume. In this spirit, Frank proposed that in a homogeneous Lennard-Jones liquid there should be dodecahedral clustering of particles, and that such clusters have a local free energy that is favorable relative to the free energies of face-centered cubic and hexagonal close packed clusters [10]. These clusters are transient and uniformly distributed. It follows that in an inhomogeneous liquid, such as an interfacial domain, there will be a concentration gradient of such clusters that is associated with, and dependent on, the density gradient in the liquid, thereby differentiating the properties of the interface from those of the bulk liquid. An alternative identification of structural motifs, whether in the homogeneous or inhomogeneous liquid, can be based on the occurrence of correlated density fluctuations. Although the conceptual descriptions of transient clusters and transient correlated fluctuations overlap, the identification protocol for the former is not based on and does not lead to an experimentally accessible observable, whereas the latter identification protocol does.

The theoretical tool most widely used for analysis of homogeneous liquid structure in terms of clusters, introduced by Steinhardt et al, involves calculating local bond order parameters using invariants defined by spherical harmonics related to the lines between the centers of particles and their neighbors [11-12]. A more recent theoretical tool used to parse the particle configuration into clusters is the Topological Cluster Classification (TCC) [13-18]. This algorithm identifies a bond network with a modified Voronoi construction (with cutoff bond length), then focuses attention on the shortest three, four and five membered rings in the bond network. Restricting attention to the equilibrium configurations of a liquid, the bond order analysis typically identifies a nonzero concentration of icosahedral clusters, with few other ordered clusters [11-12]. The TCC analysis typically identifies a broad range of ordered clusters, involving almost all of the particles in the liquid, with large concentrations of five-particle trigonal bipyramid and seven-particle pentagonal bipyramid clusters [15]. Notwithstanding exceptions like the model-based interpretation of x-ray absorption studies of liquid Cu used to imply the existence of weak order associated with existence of icosahedral clusters [16,18], these cluster analyses are not experimentally testable.

Consider, now, an inhomogeneous liquid, e.g. the liquid-X interface, where X is vapor or a solid boundary. Both the liquid-vapor interface and the liquid-solid interface exhibit spatially varying density distributions along the normal to the interface (longitudinal density distributions) with scale lengths of order of a few particle diameters. If the solid in contact with the liquid has in-plane structure, there can be local ordering in the liquid in planes parallel to the interface. Amongst the questions of interest in such cases is how the longitudinal and transverse order are coupled via the density gradient normal to the interface. Royall and coworkers have extended the TCC analysis to examine cluster populations in the liquid-vapor interfaces of a Lennard-Jones liquid and of a model of liquid Na [15]. The low temperature (near the triple point) density distributions along the normal are different for these model liquids: for the Lennard-Jones system it exhibits a monotonic decay from the bulk liquid density to the vapor density, while for the liquid Na model it exhibits density oscillations for several atomic diameters [15]. Restricting attention to the Lennard-Jones liquid-vapor interface, the calculations reported by Royall et al imply that near the triple point the dominant cluster is the trigonal bipyramid and the next most prevalent cluster is the pentagonal bipyramid [15]. The number densities of both of these clusters vary across the interface about twofold; other clusters are also present (a total of 13 types of clusters are identified), but with smaller densities [15]. For most of the types of clustering, the relative cluster densities as a function of position along the normal to the liquid-gas interface are about proportional to the overall point particle density except that the density of five-fold clustering in forms such as the pentagonal bipyramid decrease more than proportional to the decrease in overall point particle density [15]. The calculations also imply that near the triple point the five-particle plane of the pentagonal bipyramid cluster tends to lie parallel to the interface while the three-particle plane of the trigonal bipyramid cluster tends to lie perpendicular to the interface [15]. Nevertheless, it is concluded that the concentration of clusters with five-fold symmetry is suppressed in the interface. Royall et al do not exhibit calculations of the liquid structure, e.g. the pair correlation function, in planes parallel to the interface, nor do they discuss how the various clusters pack in those planes or how interaction between clusters in different planes affects the in-plane distribution [14, 15]. The latter is of considerable interest because, as discussed in Section 4 of this paper, its density dependence is coupled to the density gradient in a non-trivial fashion.

In this paper we report calculations of the aperture cross correlation function (ACCF) in planes located at different positions along the normal to the interfaces of three systems: (i) a Lennard-Jones liquid in contact with a smooth Lennard-Jones wall, (ii) a Lennard-Jones liquid in contact with the 100 face of face-centered cubic Lennard-Jones solid, and (iii) the liquid-vapor interface of a Lennard-Jones fluid. The ACCF [1], which provides a different measure of local ordering than do the bond orientation and TCC measures, is one of the variants of cross-correlation diffraction [1-9] that can uniquely identify a local structure. We focus attention on the similarities and differences between the transient ordered fluctuations derived from the ACCF analysis and the cluster concentration analysis derived from the TCC analysis, and on the consequences of the competition between the density-gradient induced ordering and the external field-induced (liquid-ordered solid) ordering.

## 2. Methods

We have simulated three systems of Lennard-Jones particles using the molecular dynamics package LAMMPS [19]. Each system had 18,000 mobile particles. The particle-particle potential was modeled to resemble Ar using the Lennard-Jones pair interaction

$$U(r) = 4\varepsilon\left[\left(\frac{\sigma}{r}\right)^{12} - \left(\frac{\sigma}{r}\right)^{6}\right] \; for \; r < r_{cut} \tag{1}$$

with $\varepsilon/k_B = 120$ K, $\sigma = 3.4$ Å, m = $6.69\times10^{-23}$ g, and a cut off radius $r_{cut} = 2.5\sigma$ [20]. The calculations were executed in the NVT ensemble for $10^6$ timesteps with 1 timestep = $(m\sigma^2/\varepsilon)^{1/2} \approx 683.4$ fs. In reduced units, the temperature was held constant at T* = $k_B T/\varepsilon = 0.786$ and the volume of the system was initialized to give a density $\rho = 0.810$ atoms/$\sigma^3$. These reduced values correspond to 94.4 K and 1.374 g/cm$^3$, respectively.

As noted before, System (i), subsequently referred to as the system with the "flat" interface, involved Lennard-Jones particles interacting with smoothed Lennard Jones planes at the +z and -z boundaries of the simulation box. The boundary-atom interaction parameters were the same as the atom-atom interaction parameters, thereby leading to the mobile particle-wall interaction $\varepsilon\left[\frac{2}{15}\left(\frac{\sigma}{z}\right)^9 - \left(\frac{\sigma}{z}\right)^3\right]$. System (ii), subsequently called the system with the "FCC" interface, had Lennard-Jones particles interacting with the 100 plane of face-centered cubic packed Lennard-Jones particles at the +z and -z boundaries of the simulation box. For these two systems, the simulation box size was 25.5436 $\sigma \times$ 25.5436 $\sigma \times$ 34.0582 $\sigma$ in the x, y, and z directions, respectively. System (iii), subsequently referred to as the system with the "L-V" interface, simulated the Lennard-Jones liquid between two regions of vapor. For the L-V system, the 18,000 particles were initialized at a density of $\rho = 0.810$ atoms/$\sigma^3$ and the simulation box, with fixed x and y lengths of 25.5436 $\sigma$, was expanded in the +z and -z directions to give a final box length of 85.1454 $\sigma$. There are Lennard-Jones vapor regions on both the +z and -z ends of the simulation box. The temperature was T* = 0.786.

In preparation for the calculation of the ACCF (see below), the calculated longitudinal density distributions in the interfaces of Systems (i) and (ii) were divided into strata with width $\Delta z$. For a stratum centered at a peak in the distribution $\Delta z$ is chosen to be the full width at half height; for a stratum centered at a trough in the distribution $\Delta z$ is chosen to be the average of the peak widths. For the L-V system, the monotonic longitudinal density profile was divided into strata with width $\Delta z = 0.5 \sigma$. To search for structured fluctuations in these strata we calculate the structure function, S(**q**), the pair correlation function, $g\left(\frac{r}{\sigma}\right)$, and the ACCF for each interface.

We have computed S(**q**) for the particle configurations derived from the simulations from

$$S(\mathbf{q}) = \frac{1}{N}\langle I(\mathbf{q},t)\rangle, \tag{2}$$

$$I(\mathbf{q},t) = \sum_{ij}^{N}\cos[\mathbf{q}\cdot\{\mathbf{r_i}(t) - \mathbf{r_j}(t)\}], \tag{3}$$

using the last 100,000 timesteps of each simulation. Particle coordinates were sampled every 1000 timesteps and converted into an image in pixels. The squared absolute value of the Fourier transform of each image was then taken to produce the corresponding instantaneous structure function, with subsequent averaging of the instantaneous structure functions as in Eq. (2). In Eq. (3) the sum is over all of the particles in a single stratum of the longitudinal density distribution. We remind the reader that S(**q**) can be experimentally determined from the intensity of radiation scattered from the system that is monitored with a single detector.

The same-time normalized ACCF of a system, defined by

$$C(\mathbf{k}, \mathbf{q}) = \frac{\langle I(\mathbf{k})I(\mathbf{q})\rangle}{\langle I(\mathbf{k})\rangle\langle I(\mathbf{q})\rangle}, \tag{4}$$

provides information not contained in S(**q**). In Eq. (4), the average is over particles contained within a small aperture in a single stratum of the longitudinal density profile. Unlike S(**q**), to measure C(**k**,**q**) the scattered intensity is measured with two detectors located to monitor scattered radiation with wave vectors **q** and **k**. When one detector is kept stationary and |**k**| = |**q**|, the cross correlated intensity with respect to angle takes the form

$$C(\varphi) = \frac{\langle I(\varphi_0)I(\varphi_0+\varphi)\rangle}{\langle I(\varphi_0)\rangle\langle I(\varphi_0+\varphi)\rangle} \tag{5}$$

with $\varphi_0$ constant corresponding to the fixed direction of the detector at **k** and ($\varphi_0 + \varphi$) the angle between the detectors at **k** and **q**. We note in passing that the principal difficulty in making ACCF measurements in experimental situations is screening out unwanted scattering from the region outside the aperture chosen, which can be readily achieved in two dimensions and with considerable effort in three dimensions. Choosing |**k**| = |**q**| implies that the structure of an ordered fluctuation can be inferred from the intensity distribution in one ring of the diffraction pattern. C(**k**, **q**) can also be calculated using different values of **k** and **q** when that is required to identify the local structure. In the current study, the radius of the circular aperture was 4σ. Small, circular areas of the cross section were selected from the trajectory file until every area of the stratum had been sampled at least once. Both the pair correlation functions and the aperture cross correlation functions in this study were calculated over the last 900,000 timesteps of the simulations, when the systems were equilibrated

## 3. Results

### A. Longitudinal Density Distributions and

Figure 1 depicts the longitudinal density profiles of the three systems studied with the z widths of the strata representing cross-sections at the peaks delineated by colored bars. The longitudinal density distributions in the Lennard-Jones liquids in contact with both the flat and the 100 face of FCC Lennard-Jones walls display pronounced oscillations extending about 5σ into the liquid phase (Fig. 1). The ratio of first peak-to-first trough point densities is about 3 for liquid in contact with the flat wall, and about 13 for liquid in contact with the FCC wall. In both cases the peak to trough point density ratio decreases smoothly with increasing distance from the wall, reaching unity at about $6\sigma$. In contrast, the longitudinal density profile in the liquid-vapor interface of the Lennard-Jones liquid, at the same temperature and liquid density, is a monotonic decaying function spanning about $4\sigma$.

**B. Transverse Structure Functions**

Homogeneous liquids with densities as different as at the values cited above for the first peak and trough of the liquids in contact with either the FCC or flat wall have pair correlation functions that are distinctively different. That is not the case for the inhomogeneous liquids we examine in this paper. For each case we have considered, the stratum in the longitudinal density profile that is furthest from the interface is sensibly bulk liquid. Consider now the liquid-flat wall and liquid-vapor interfaces. Examination of the pair correlation functions calculated for strata that represent cuts in the liquid-flat wall longitudinal distribution function at the peaks and troughs or in the monotonic liquid-vapor longitudinal distribution function (Figs. 2 - 6) and the corresponding diffraction patterns (Figs. 7 and 8) show that the pair correlation functions in the several liquid strata in these inhomogeneous interfaces are sensibly identical, i.e., independent of the nominal local point densities of the longitudinal density distribution. In contrast, the pair correlation functions of the liquid in the two strata closest to the FCC wall display structure not found in the liquid (Figs. 4 and 5). This structure disappears in strata further away from the interface, and in those strata the pair correlation functions are sensibly identical with the bulk liquid pair correlation function. The structure functions of the strata in the liquid-FCC interface suggest FCC structure among the mobile argon atoms two layers within the liquid (Fig. 7). Even the third layer displays some four-fold preference in the first diffraction ring. The absence of such strong four-fold peaks in the later layers of the liquid-FCC interface implies that atoms at these longitudinal levels do not favor locally arranging in transient FCC-like structures.

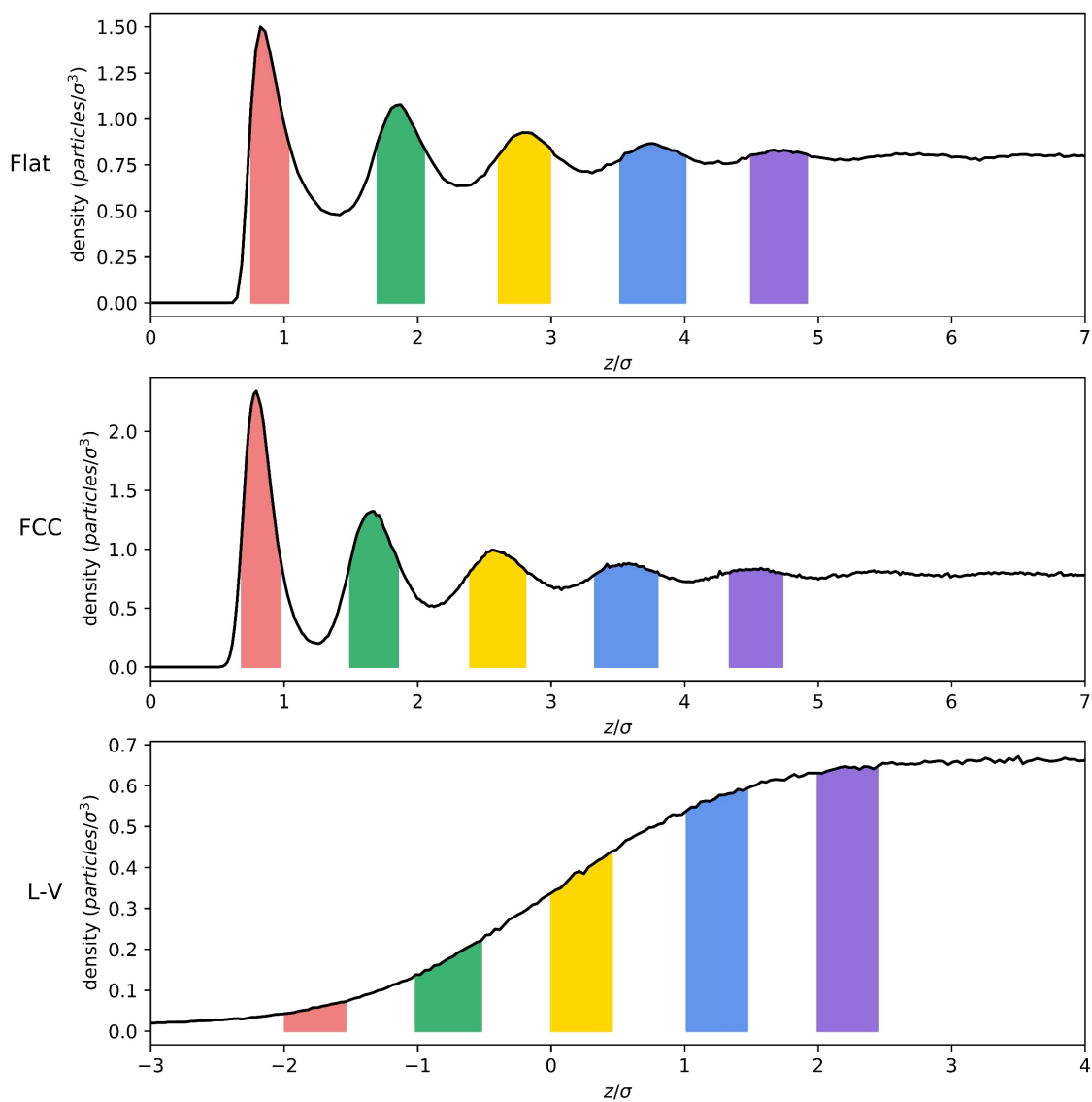

**Figure 1**. Longitudinal density profiles of Lennard-Jones particles in the interfaces between liquid and (A) a flat Lennard-Jones boundary, (B) the 100 face of a FCC Lennard-Jones boundary, and (C) vapor. These profiles are averaged over the z+ and z- interfaces. For panels A and B the boundary wall is located at $z/\sigma = 0$; for panel C, the half-density location is selected as the origin. The colored bars indicate the range of z values that each cross-section spans. $T^* = 0.786$. Note that the vertical scales of A, B and C are different.

**C. Aperture Cross Correlation Functions**

The ACCFs (Figs. 9 - 14) provide information that both supplements and is not obtainable from the structure functions and the longitudinal density distributions. Consider, first, the liquid-FCC wall interface (Fig. 9). There is a very strong signature of four-fold structure in the first two strata, weak signatures of that structure in strata three and four, reverting to disorder (no peaks) in the fifth stratum. This structure is imposed on the liquid by the atomic arrangement of the FCC 100 face. Note that no such structure appears in the trough regions between the strata (Fig. 10). Furthermore, the ACCFs of the liquid-flat wall interface and the liquid-vapor interface (Figs. 11-12) do not display a preference for ordered fluctuations with specific symmetries.

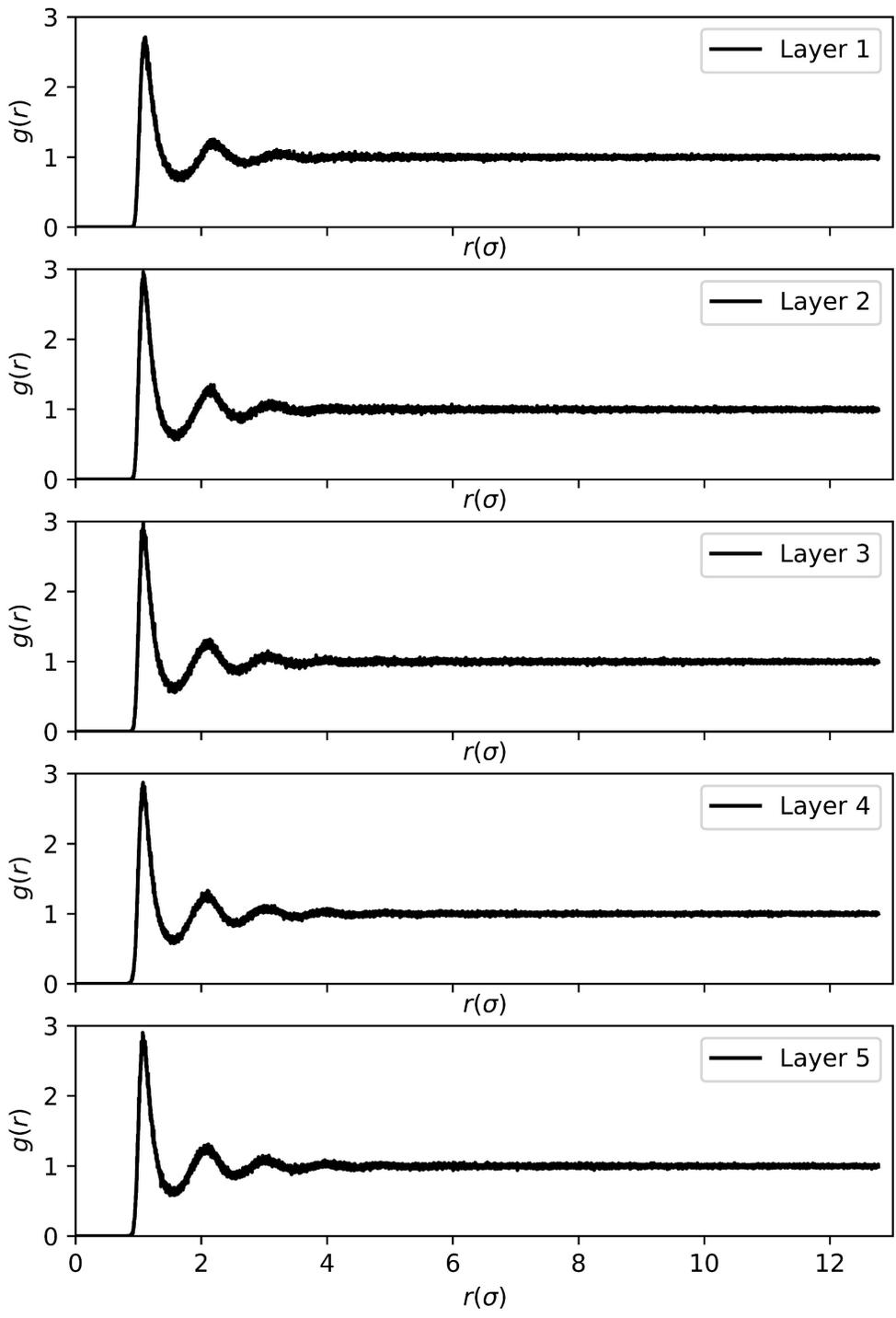

**Figure 2.** 2D pair correlation functions for several strata at the peaks in the interfacial longitudinal density distribution of liquid Lennard-Jones particles interacting with a flat Lennard-Jones wall. Layer 1 is the stratum closest to the wall.

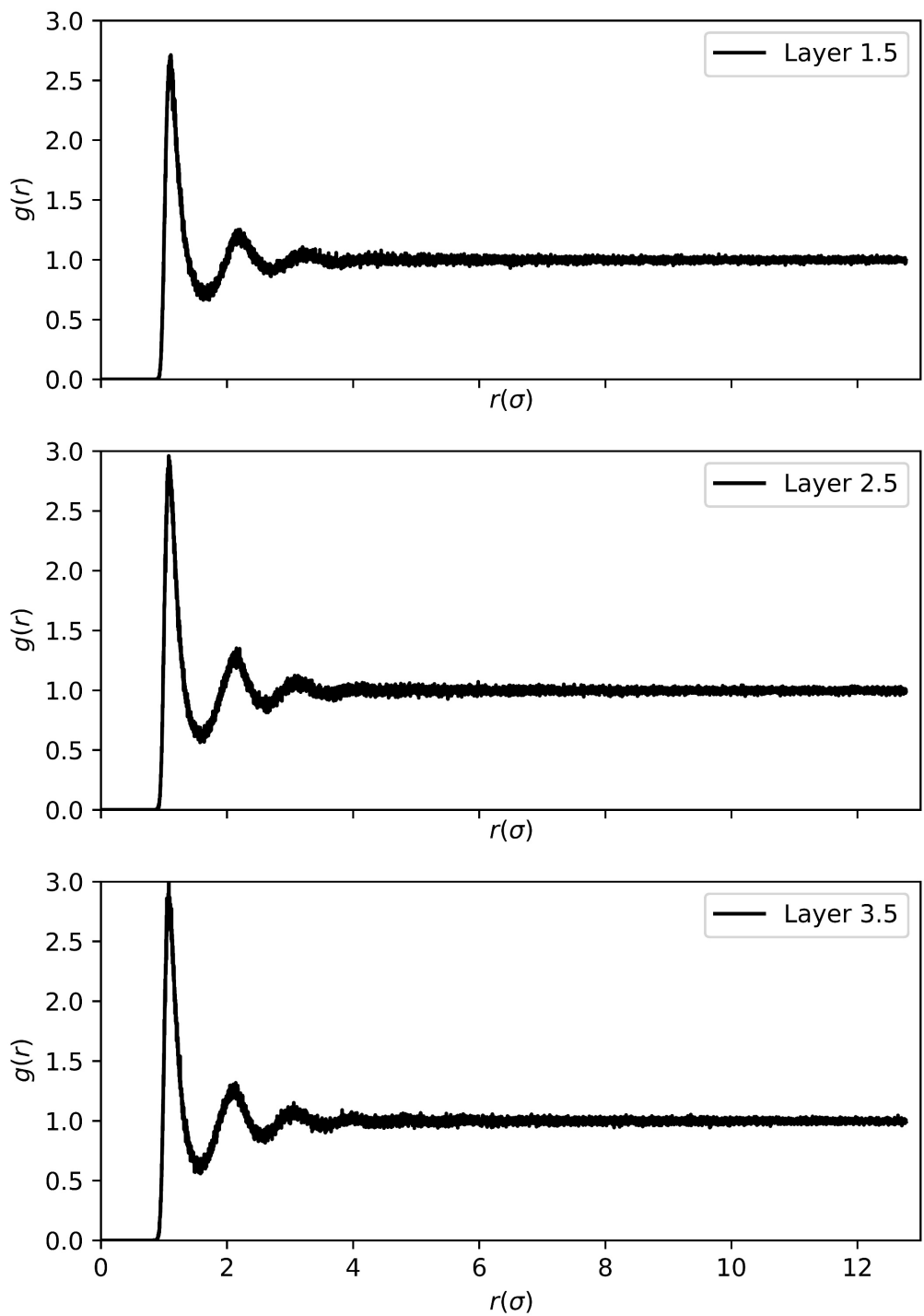

**Figure 3.** 2D pair correlation functions for several strata at the troughs in the interfacial longitudinal density distribution of liquid Lennard-Jones particles interacting with a flat Lennard-Jones wall.

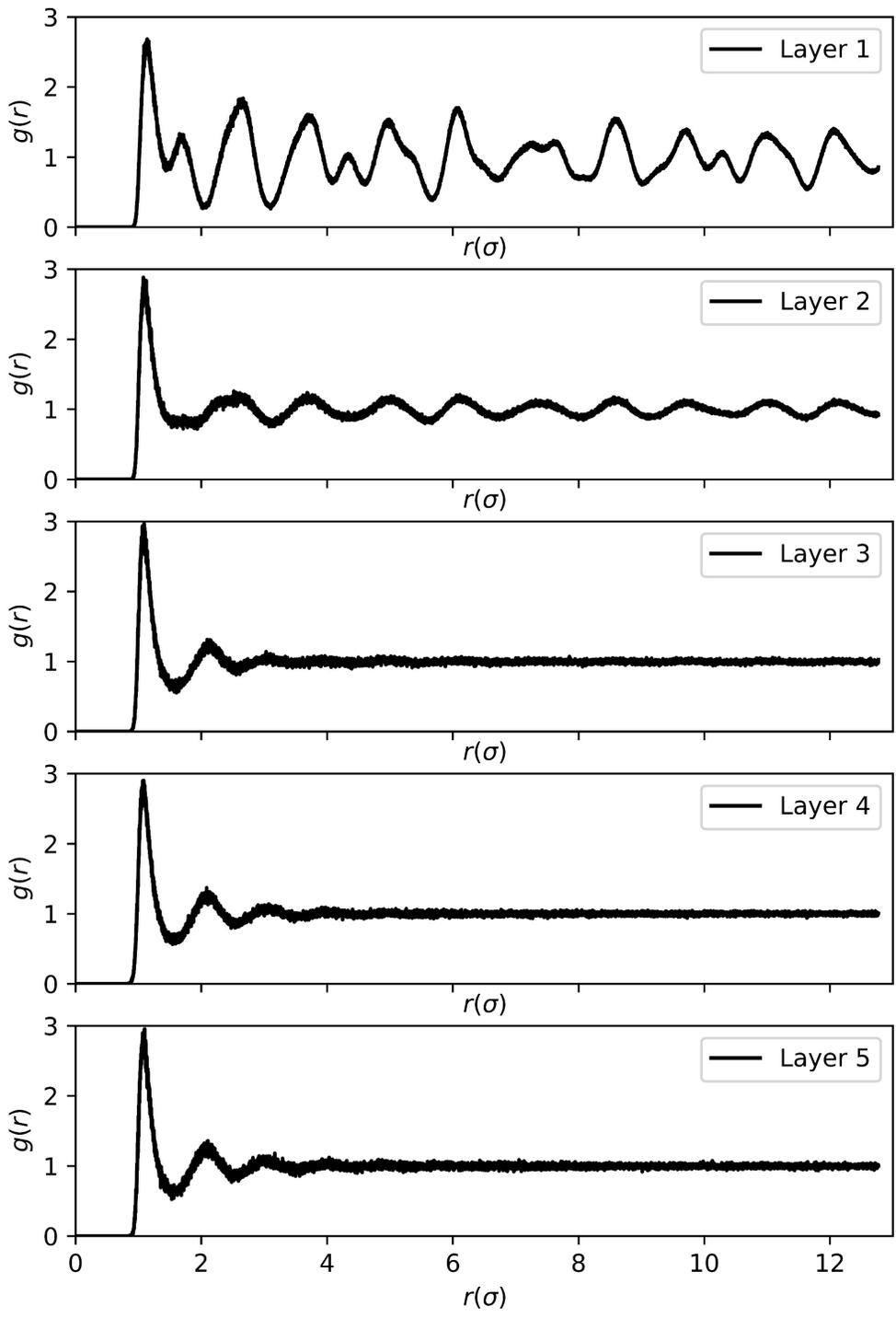

**Figure 4.** 2D pair correlation functions for several strata in the interfacial longitudinal density distribution of liquid Lennard-Jones particles interacting with the 100 plane of a FCC Lennard-Jones wall. Layer 1 is the stratum closest to the wall.

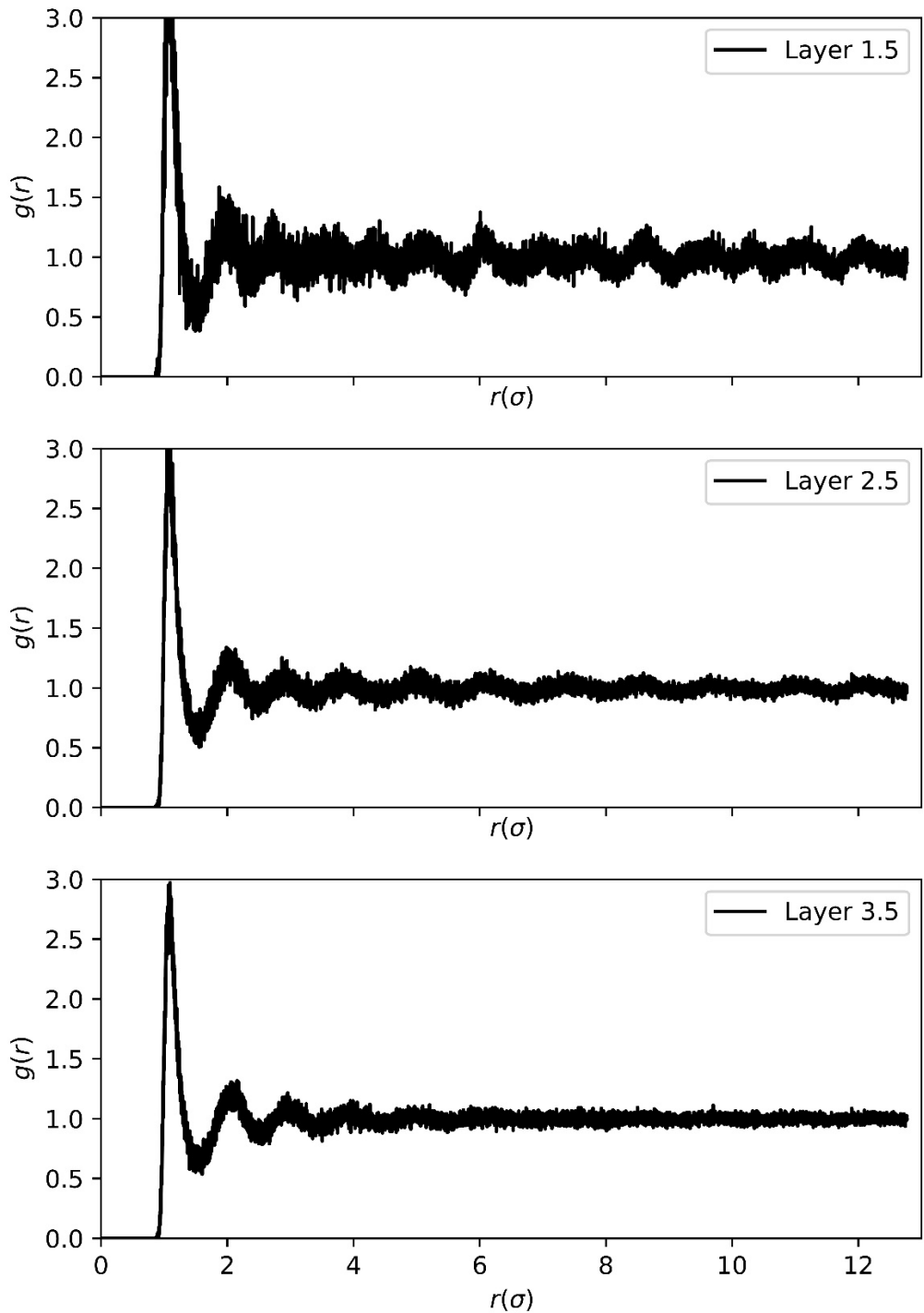

**Figure 5.** 2D pair correlation functions for several strata at the troughs in the interfacial longitudinal density distribution of liquid Lennard-Jones particles interacting with the 100 plane of a FCC Lennard-Jones wall.

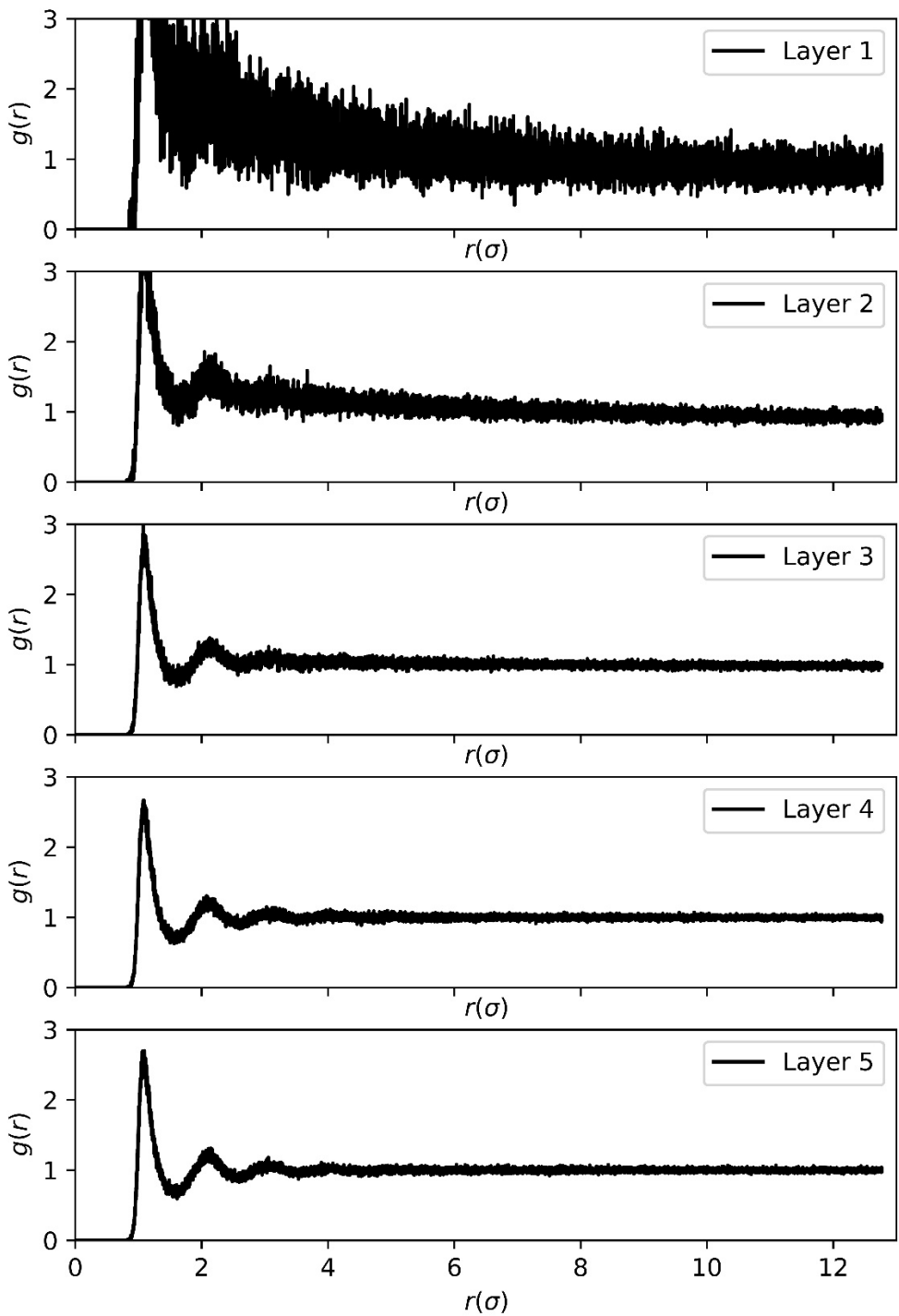

**Figure 6.** 2D pair correlation functions for several strata in the longitudinal density distribution of the liquid-vapor interface of liquid Lennard-Jones particles. Layer 1 is the closest to the vapor phase.

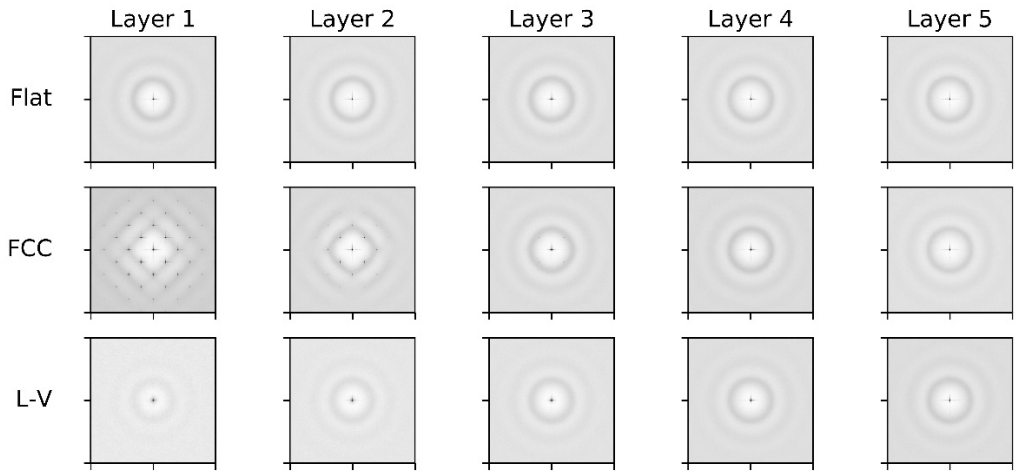

**Figure 7**. Diffraction patterns (structure functions) of simulated Lennard-Jones liquid in strata located at the peaks of the longitudinal density distributions of (A) System #1, (B) system #2, and (C) System #3, the liquid-vapor interface. In panels A and B, layer 1 is the closest to the boundary. In panel C, layer 1 is the closest to the vapor phase. T* = 0.786.

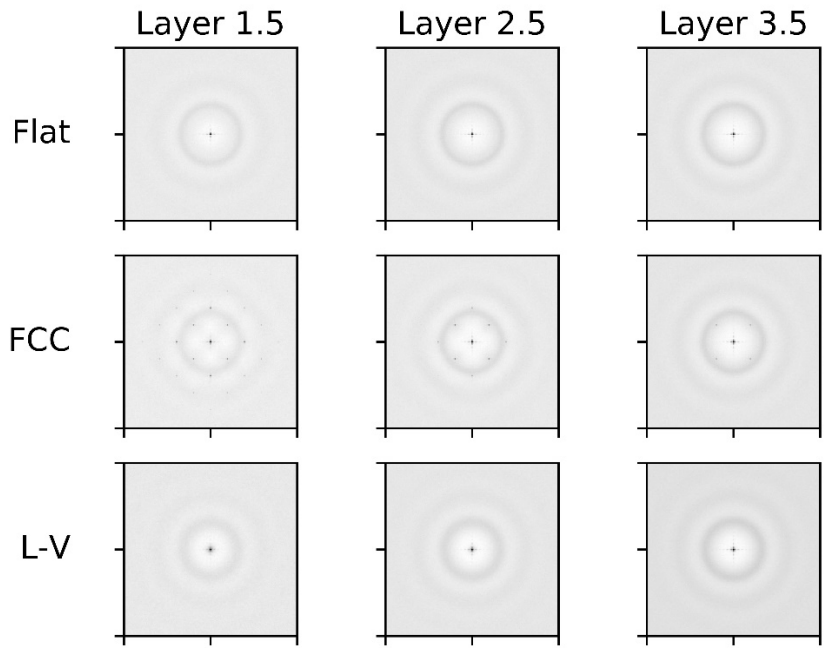

**Figure 8**. Diffraction patterns (structure functions) of simulated Lennard-Jones liquid in strata located at the troughs of the longitudinal density distributions of (A) System #1, (B) system #2, and (C) System #3, the liquid-vapor interface. In panels A and B, layer 1 is the closest to the boundary. In panel C, layer 1 is the closest to the vapor phase. T* = 0.786.

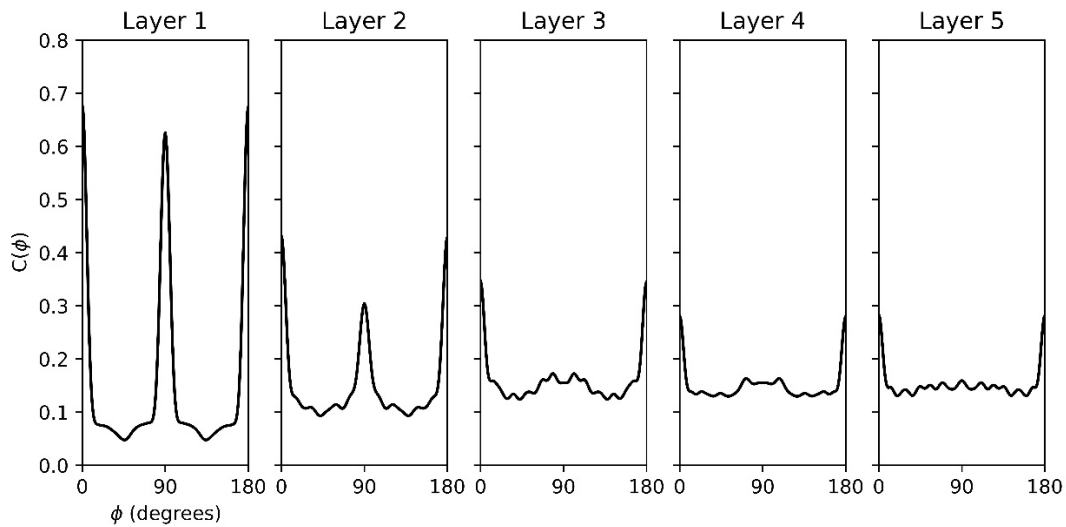

**Figure 9.** Aperture cross-correlation functions of simulated Lennard-Jones liquid interacting with the 100 face of a FCC lattice of Lennard-Jones particles. These functions are produced along the first diffraction ring. Layer 1 is the cross section closest to the lattice.

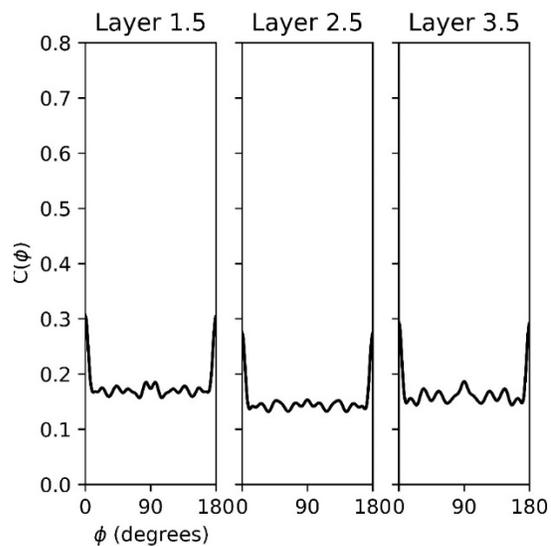

**Figure 10.** Aperture cross-correlation functions of simulated liquid Lennard-Jones liquid interacting with the 100 face of a FCC lattice of Lennard-Jones particles. These functions are produced along the first diffraction ring. Layer 1.5 is the cross section closest to the lattice interface.

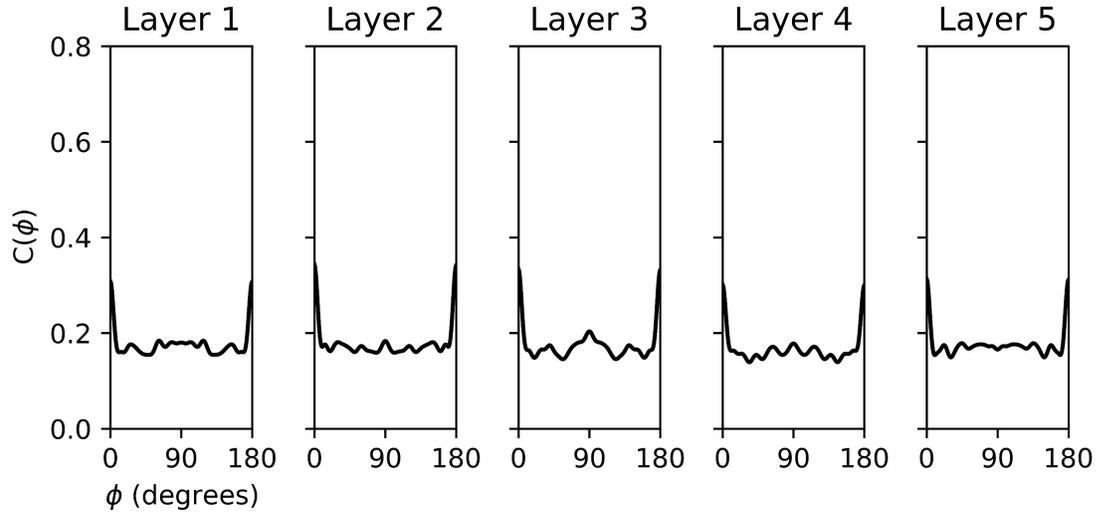

**Figure 11.** Aperture cross-correlation functions of simulated Lennard-Jones liquid interacting with a flat Lennard-Jones wall. These functions are produced along the first diffraction ring. Layer 1 is the cross section closest to the Lennard-Jones wall.

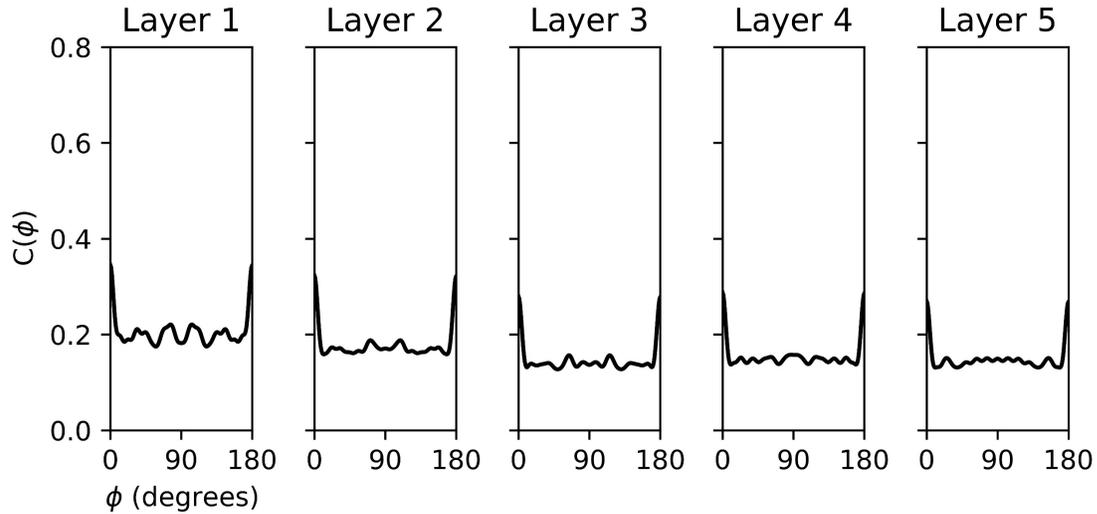

**Figure 12.** Aperture cross-correlation functions of a simulated Lennard-Jones liquid-vapor interface. These functions are produced along the first diffraction ring. Layer 1 is the closest to the vapor phase.

## 4. Discussion

We consider first the strata and density dependences of the transverse pair correlation functions in the several systems studied. The results of the simulations show clearly that, except for the first two strata in the liquid-FCC wall interface and the outermost least dense stratum of the liquid-vapor interface, the pair correlation functions are sensibly identical in the several strata in all of the studied inhomogeneous interfaces, i.e., independent of the local point densities of the longitudinal density distribution. In contrast, the pair correlation functions of the liquid in the two strata closest to the FCC wall display structure not found in the liquid.

Taking the latter observation first, the observed behavior of the pair correlation function is an example of the modulation of the structure of a 2D liquid induced by a periodic field. Moss et al [21] provide a full description of this phenomenon via analysis of x-ray scattering from a monolayer of liquid Rb inserted between two graphite planes. The theoretical analysis of the x-ray scattering from a 2D liquid, developed by Reiter and Moss [22], predicts both modulation of the liquid structure function with the period of the applied field and replication of the liquid structure function about the reciprocal lattice points of the applied periodic field. The pair correlation functions for layers 1 and 2 in Fig. 4 show the first effect, and the corresponding diffraction patterns and ACCFs in Figs. 7 and 9 show the square symmetry of the 100 face of the FCC lattice imposed on the liquid.

Consider, now, the former observation. We interpret the lack of dependence of the transverse pair correlation functions on the point densities of the longitudinal density distribution to be a fundamental consequence of the delocalization of the thermodynamic force that maintains the longitudinal density gradient. The first equation of the Bogoliubov-Born-Green-Kirkwood-Yvon hierarchy, for the spatial dependence of the single particle distribution function, represents the condition of mechanical equilibrium in a liquid subject to an external force. That external force generates the density gradient in the system or, put another way, the thermodynamic force generated by the density gradient balances the external force. The conventional form of the local equilibrium assumption states that the properties of an inhomogeneous liquid at some point are the same as those of a homogeneous liquid with the same point density. But in a system of molecules, the thermodynamic force associated with the density gradient is generated by interactions between the molecules, hence must be delocalized over a volume no smaller than a sphere with radius equal to the molecular diameter. The essence of the Fischer-Methfessel approximation is replacement of the point density definition of local equilibrium with a definition that uses the point density averaged over a small volume centered between a pair of molecules. Applying the lowest order approximation to calculate this average (i.e. an average over the sphere with radius equal to the molecular diameter) from our longitudinal density distributions, it is easily seen that the average of the point density over the peaks and troughs is sensibly equal to the bulk liquid density. That being the case, the transverse pair correlation function is expected to be independent of the local point density. A detailed analysis of just this type [23], for the stratified liquid-vapor interface of Cs, including least squares fits to determine the density of the homogeneous liquid with pair correlation function that best matches that of the inhomogeneous liquid at a selected point on the longitudinal density profile, yields the same result as cited for the data presented in this paper.

As noted in Section 3, the ACCFs of the liquid-flat wall interface and the ACCFs of the liquid-vapor interface do not reveal preference for structured fluctuations with a specific symmetry. While there appear to be slightly larger peaks in the ACCFs of the outermost layers of both interfaces, further analysis using averages over separated time segments and additional simulations show that these peaks are within the noise level of the calculations. Studies of a 2D liquid and of a liquid closely confined (say up to five molecular diameters) between parallel walls have shown that such liquids can support transient structured fluctuations with hexagonal and square symmetries [24-26]. Therefore, we suggest, from the absence of ACCF peaks in the stratum closest to the Lennard-Jones wall, that sufficient free volume for out-of-plane motion lowers the stability of any favored in-plane ordered structures. The same observation and argument hold for the liquid-vapor interface.

To establish a connection between the representations of local order using the ACCF and the TCC analysis we turn to the study of the Na and the Lennard-Jones liquid-vapor interfaces reported by Royall et al [15]. We consider, first, the liquid-vapor interface of Na that they model using a potential, due to Chacon, that generates a large ratio of critical to triple point temperatures and a stratified longitudinal density profile. We take this calculation to provide an analogue to the interfacial profile we calculate for a Lennard-Jones-flat wall system, noting that the longitudinal density modulations in the latter are larger than in the former. After parsing the particle configurations into multiple (13) cluster types, Royall et al find that in the low temperature bulk liquid the most popular clusters, including the pentagonal bipyramid, account for more than half the particles [15]. They also find that the relative densities of all of the cluster types decrease as the interface is traversed towards the vapor, with a slightly faster decrease of clusters with five-fold symmetry, so that there is no enhancement at the surface, relative to the bulk, of clusters with five-fold symmetry. With regard to that enhancement, Royall et al report that the pentagonal planes of the pentagonal bipyramid clusters show a slight preference to be oriented parallel to the interface, whereas the triangular planes of the trigonal bipyramid clusters show a slight preference to be oriented perpendicular to the interface. To the extent that a comparison of this model of the liquid-vapor interface of Na and our Lennard-Jones liquid-flat wall model is valid, the descriptions of the local structure by the TCC and ACCF analyses do not mesh. Taken at face value, the TCC prediction of a very low concentration of clusters with five-fold symmetry in the outermost layer of the interface which increases as one goes deeper into the liquid is not supported by the lack of dependence of the ACCFs on depth in the interface. As to comparison of the predicted local structures in the Lennard-Jones liquid-vapor interface, Royall et al find the same general trends for cluster concentration as for Na as the interface is traversed towards the vapor [15]. Finally, the reported TCC cluster analysis of the liquid-vapor interface does not include any reconstruction from the cluster species and concentrations of the transverse pair correlation function. It is not obvious how, given the calculated variation of cluster species and concentrations with position in the interface, such a reconstruction will generate pair correlations that do not depend on position along the density profile in the interface.

Considering the arguments advanced above, we interpret the ACCF and TCC descriptions of local ordering in an inhomogeneous liquid to be extremes in the representation of the local structure of inhomogeneous liquid. In particular, the TCC description implicitly assigns all of the particles to clusters, so the liquid is at all times a mixture of transient ordered species with well-defined average concentrations. In contrast, the ACCF description deals directly with the equilibrium particle configurations, whether experimentally generated or obtained from a simulation, and identifies the local structures that arise from thermally induced correlated density fluctuations in the inhomogeneous system. In the TCC analysis density fluctuations must be constructed from concentration fluctuations of the several species of clusters. The locally ordered transient fluctuations identified by an ACCF analysis are not clusters as defined by the TCC analysis. The ACCF analysis is designed to generate an experimental observable with different signatures for different transient ordered fluctuations. As such, it does not depend on a particular definition of local equilibrium in the inhomogeneous liquid, and it does not depend on prior identification of the symmetries of the correlated fluctuations. In contrast, the TCC analysis depends on prior identification of the cluster species to be found and on the point density definition of local equilibrium in the inhomogeneous liquid. Just as the use of separate expansions of a function in terms of different complete sets of orthonormal functions can yield, by comparison, insights not available from one expansion, so can comparison of ACCF and TCC (or bond order) descriptions of transient local order in an inhomogeneous liquid.

## 5. Acknowledgements

The data that supports the findings of this study are available within the article.

This research was funded by the University of Chicago Materials Research Science and Engineering Center and the National Science Foundation (Grant No. DMR-1420709).